# Clausius-Mossotti approximation in the theory of polar materials

*Y. Kornyushin*

Maître Jean Brunschvig Research Unit
Chalet Shalva, Randogne, CH-3975, Switzerland
email: jacqie@bluewin.ch

Clausius-Mossotti approximation is extended to describe the measured magnetic moment of an ellipsoidal sample containing magnetic or nonmagnetic ellipsoidal inclusions and magnetic or nonmagnetic matrix. The magnetic field in the matrix and inclusions is calculated. The magnetic energy of a system is calculated also. The equilibrium shape of a pore in a ferromagnetic sample is investigated. The phenomenon of cavitation in porous ferromagnetic samples is described. The model is applied to calculate magnetic properties of granular superconductors. The effective electric conductivity of a sample of a composite material, containing an arbitrary number of differently ordered distributions of ellipsoidal inclusions is calculated. Effective conductivity of a composite material, consisting of fibers of high conductivity and a matrix of low conductivity is discussed. Concentrated electric field in the vicinity of the ends of a conductive nanofiber in a composite material is calculated. The high quantity of this field is of an extreme importance to provide the proper functioning of monitors, based on conductive nanofibers in dielectric media.

**1. Introduction**

Clausius-Mossotti approximation, which was introduced by Ottavanio Fabrizio Mossotti in 1846, is now one of the models used to describe the effective conductivity or susceptibility of mixtures and materials containing several phases [1]. Problems of the effective susceptibility and conductivity of inhomogeneous samples are identical from the point of view of the mathematical approach. It is easy to see why it is so. The magnetic inductance $\mathbf{B}(\mathbf{r}) = \mathbf{H}(\mathbf{r}) + 4\pi\mathbf{M}(\mathbf{r})$ ($\mathbf{H}$ and $\mathbf{M}$ are the magnetic field and magnetization, correspondingly) is related to the magnetic field by the following equation:

$$\mathbf{B}(\mathbf{r}) = (1 + 4\pi\chi)\mathbf{H}(\mathbf{r}), \qquad (1)$$

where $\chi$ is the susceptibility.

For an inhomogeneous medium $\chi$ depends on the coordinates $\mathbf{r}$ and the local values of $\mathbf{H}(\mathbf{r})$ are determined by the continuity equation, $\text{div}\mathbf{B}(\mathbf{r}) = 0$. The effective value of $\chi$, $\chi_e$, is determined by the following equation,

$$\langle\mathbf{B}\rangle = (1 + 4\pi\chi_e)\langle\mathbf{H}\rangle, \qquad (2)$$

where $\langle ... \rangle$ denotes averaging over the volume of a sample.

Equations (1) and (2) could be applied as well for the problem of electric conductivity, only we have to substitute the current density $\mathbf{J}(\mathbf{r})$ instead of $\mathbf{B}(\mathbf{r})$ and the electric field $\mathbf{E}(\mathbf{r})$ instead of $\mathbf{H}(\mathbf{r})$, and the electric conductivity $\sigma(\mathbf{r})$ instead of $(1 + 4\pi\chi)$. The problem of the calculation of the effective conductivity (susceptibility) of inhomogeneous media has been considered by many authors, see e.g. [1 – 4]. The simplest approach is to approximate $\mathbf{E}(\mathbf{r})$ by $\langle \mathbf{E} \rangle$. This immediately yields $\sigma_e = \langle \sigma \rangle$. Another approach is to approximate $\mathbf{J}(\mathbf{r})$ by $\langle \mathbf{J} \rangle$. This yields $\sigma_e = \langle \sigma^{-1} \rangle^{-1}$. Then it was shown that the actual value of $\sigma_e$ always lies between $\langle \sigma^{-1} \rangle^{-1}$ and $\langle \sigma \rangle$: $\langle \sigma^{-1} \rangle^{-1} \leq \sigma_e \leq \langle \sigma \rangle$ [4].

In this paper we shall consider the case of a sample in an external applied field, whereas the sample consists of several phases in a matrix. We shall use the approximation of the field (induced by the external one) having constant (but different) values in each phase and in the matrix of a sample. Solution of the problem in such an approximation may be obtained for the problem of susceptibility, in particular. After obtaining a solution for the susceptibility we shall return to the problem of the effective conductivity.

In an ellipsoidal sample, which consists of ellipsoidal magnetic particles in a nonmagnetic matrix, the field in the matrix is not really the homogeneous one as every particle is a dipole. The distribution of the field in the matrix volume of such a sample is considered in this paper. The distribution of the field in the volume of a matrix is shown to be nonsymmetrical and the first three moments of the field distribution are calculated. The cases of the equidistant and the random distributions of the particles in a matrix are considered separately. The shift of the volume-averaged field respectively to the external applied one is calculated.

The most general solution of Clausius-Mossotti approximation is obtained. The case of the ellipsoidal magnetic sample, containing ellipsoidal magnetic inclusions with different values of the magnetization is considered. The inclusions are assumed to be of different types (the magnetizations, the volumes, and the depolarization factors (see **Appendix C**) are different for the different types of the inclusions), but it is assumed that all the inclusions are oriented along the external applied field $\mathbf{H}_0$. In a general case the external applied field is not oriented along one of the axes of the inclusion and obtained equations refer to a corresponding components of the vector values considered.

In the condition, when the magnetic field penetrates through the volume of a ceramic superconductive sample, the value and the variation of the field inside a superconductive sample and measured magnetic and superconductive properties attract attention of the authors of some resent publications [5 – 7]. An application of the theory to the cases of ceramic superconductors is given below too. This simple theory allows understanding changes, occurring in the weak links (see **Appendix C**) with the increase of the field.

Obtained results are important for a description of the magnetic and other physical properties of sintered materials from ceramic superconductors to porous magnetic materials with growing pores and first order ferromagnetic phase transformations.

## 2. Magnetic inclusions in a nonmagnetic matrix

Let us regard an ellipsoidal sample, with the depolarization factor of a sample as a whole $N$, which consists of an arbitrary number of magnetic phases, immersed in a nonmagnetic matrix [8]. Let the volume fractions of the magnetic phases in the sample be $f_k$, the total fraction of the magnetic material $f = \Sigma f_k$, and the matrix volume fraction $1 - f$. The value of the external applied field is $H_0$, the value of the field in the matrix is $H_m$, the magnetization of the $k$-th phase is $M_k$, and the depolarization factors of the grains of the $k$-th phase are $n_k$. All the fields and the magnetizations are assumed to be parallel to the external field $\mathbf{H}_0$. For such system of orderly oriented magnetic grains of arbitrary number of phases it is possible to formulate a model and to obtain an approximate solution of the problem. As $H_m$ serves as the external field in respect to the magnetic grains, the value of the magnetic field inside a grain in such a model is assumed to be [2]

$$H_k = H_m - 4\pi n_k M_k, \tag{3}$$

where $H_k$ is the magnetic field inside the $k$-th phase.

It is assumed in Eq. (3) that the boundaries between a non-magnetic matrix and magnetic grains are sharp and we have assumed also that the situation could be described by the definite effective values of the field inside a matrix as well as inside grains of each phase. And now we shall finally assume that the averaged over the volume of a sample value of the magnetic field inside a sample $\langle H \rangle$ is determined by the averaged over the sample volume magnetization $\langle M \rangle$:

$$\langle H \rangle = H_0 - 4\pi N \langle M \rangle = H_0 - 4\pi N \Sigma f_k M_k. \tag{4}$$

On the other hand, direct averaging of the field inside a sample yields:

$$\langle H \rangle = (1 - f)H_m + \Sigma f_k(H_m - 4\pi n_k M_k) = H_m - 4\pi \Sigma n_k f_k M_k. \tag{5}$$

Comparing Eqs. (4) and (5) yields [8]:

$$H_m = H_0 - 4\pi \Sigma (N - n_k) f_k M_k, \tag{6}$$

and Eq. (3) acquires the following form:

$$H_k = H_0 - 4\pi n_k M_k - 4\pi \Sigma (N - n_j) f_j M_j. \tag{7}$$

Now let us introduce the susceptibilities of the phases and the matrix:

$$\chi_k = M_k/H_k, \; \chi_m = 0. \tag{8}$$

From Eqs. (7) and (8) it is easy to obtain:

$$[\chi_k/(1 + 4\pi n_k \chi_k)]H_0 = M_k + [4\pi \chi_k/(1 + 4\pi n_k \chi_k)]\Sigma(N - n_j) f_j M_j. \tag{9}$$

Equation (9) is a system of linear equations for $M_k$. This system determines $M_k$ and $\langle M \rangle = \Sigma f_k M_k$ as functions of $H_0$, that is the effective susceptibility of a sample $\chi_e$.

## 3. The grains of the same shape, the concept of the effective depolarization factor

Now let us consider the case when $n_k = n$, i.e. the same depolarization factor for all the magnetic particles in a sample. For this case we shall readily get:

$$\langle M \rangle = H_0[\Sigma f_k \chi_k/(1 + 4\pi n \chi_k)]/[1 + 4\pi(N - n)\Sigma f_k \chi_k/(1 + 4\pi n \chi_k)]. \tag{10}$$

Let us regard now a sample, which consists of a single magnetic phase in a non-magnetic matrix (one of the possible structures of the sample is shown in Fig. 1). The external magnetic field $\mathbf{H}_0$ is directed along one of the main axes of the ellipsoidal sample and of the grains.

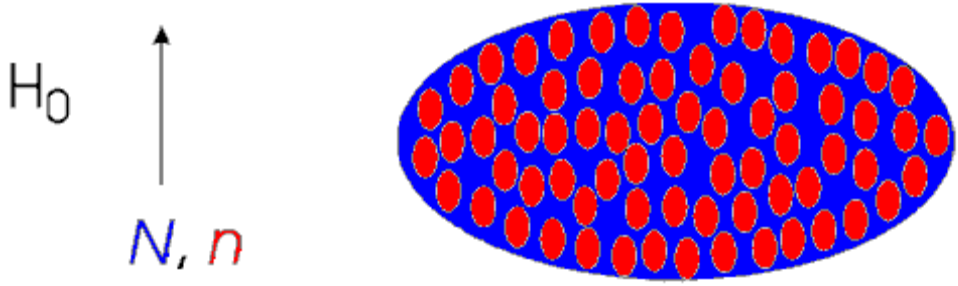

**Fig. 1.** Extension of Clausius-Mossotti model for granular magnetic materials. Here $\mathbf{H}_0$ is the external magnetic field, $N$ is the depolarization factor of a sample as a whole, and $n$ is the depolarization factor of grains.

For the case of a single magnetic phase with magnetization $M$ we have:

$$\langle M \rangle = f\chi H_0 / \{1 + 4\pi\chi[fN + (1-f)n]\}. \tag{11}$$

Comparing this equation to a well-known relationship for a homogeneous sample,

$$\langle M \rangle = \chi H_0 / (1 + 4\pi\chi N), \tag{12}$$

we conclude, that the effective depolarization factor of a granular sample is given by the following expression:

$$N_e = fN + (1-f)n. \tag{13}$$

This result was obtained in [8].

For the case of completely shielded superconductive grains $\mathbf{B} = 0$ inside the grains and, as follows from Eq. (1), $\chi = -1/4\pi$. The case of the partial penetration of the field through the superconductive grains may be described by the value of $\chi$ different from $-1/4\pi$, or by introducing magnetic induction, penetrating superconductive grains $B_s$ and, as it is shown in [9], in this case

$$\langle M \rangle = f(B_s - H_0)/4\pi(1 - N_e). \tag{14}$$

Equation (14) will be derived in this paper in **Section 13**.

Equations (13) and (14) were used in [9] to treat the experimental data, obtained on the granular samples of superconductive $YBa_2Cu_3O_7$. This treatment allowed to estimate the lower and the higher critical fields $H_{c1}$ and $H_{c2}$ for weak links, $H_{c1}$ for the grains, and to evaluate $f$ for different samples. Obtained results are in a good agreement with previously published ones and data, obtained by independent measurements. These data will be considered in detail later, they are for the favor of the presented simple model.

It is worthwhile to note that the quantities $B_s$ and $\chi$ are interdependent, because as follows from Eq. (11) and relations $\langle M \rangle = fM$ and $M = \chi H_s$ ($H_s$ is the magnetic field inside a superconductive grain),

$$B_s = H_0(1 + 4\pi\chi)/(1 + 4\pi\chi N_e). \tag{15}$$

Equation (15) allows obtaining the value of $\chi$ if the value of $B_s$ is known and vice versa.

**4. Two completely shielded superconductive phases in a non-superconductive matrix**

For this case we have:

$$\langle M \rangle = -(H_0/4\pi)[f_1(1 - n_2) + f_2(1 - n_1)]/$$

$$[(1 - n_1)(1 - n_2) - f_1(N - n_1)(1 - n_2) - f_2(N - n_2)(1 - n_1)]. \qquad (16)$$

Often it is important to know the value of the field in the matrix (it determines a behavior of the weak links). This value is presented by the following relationship:

$$H_m = H_0(1 - n_2)(1 - n_1)/[(1 - n_1)(1 - n_2) - f_1(N - n_1)(1 - n_2) - f_2(N - n_2)(1 - n_1)]. \qquad (17)$$

Equation (3) may be used to calculate the $H_k$ values. For the cubically symmetric in average case, when $f_1 = f/3$, $f_2 = 2f/3$, $n_1 = n$, $n_2 = (1 - n)/2$, we have:

$$\langle M \rangle = \langle (H_0/4\pi)f(5 - 3n)/[3(1 - n^2) + f(2 + 3n^2 - 3n + 3Nn - 5N)], \qquad (18)$$

and for the field in the matrix we have:

$$H_m = 3H_0(1 - n^2)/[3(1 - n^2) + f(2 + 3n^2 - 3n + 3Nn - 5N)]. \qquad (19)$$

These equations may be used for the calculation of the values of the magnetic moment and the field in the weak links and in the matrix of a sample.

Now let us consider a numerical example. Let us calculate a field in the matrix for different directions of the external applied field. For the ellipsoid of revolution with $N = 0.53$, $f = 0.7$, and $n = 0.6$ we have $H_{mc}/H_{mr} = 1.406$ ($c$ refers to the direction of the applied field parallel to the short axis, $r$ - parallel to the long axis). For $N = 0.53$, $f = 0.7$, and $n = 0.25$ we have $H_{mc}/H_{mr} = 1.562$. It is interesting to compare these values to the case of spherical grains, $n = 1/3$. For $N = 0.53$, and $f = 0.7$ we have $H_{mc}/H_{mr} = 1.39$. So the difference between the three cases is no more than 11% only.

## 5. Application to conductivity

To obtain formula for the effective conductivity $\sigma_e$ it is necessary to write down the relationship for $\langle B \rangle = H_0 + 4\pi(1 - N)\langle M \rangle$ and to express it through $\langle H \rangle = H_0 - 4\pi N \langle M \rangle$, according to Eq. (2), and then to replace $(1 + 4\pi\chi_k)$ by $\sigma_k/\sigma_m$. After doing so we shall have for the case of the same shape of the grains:

$$\sigma_e = \sigma_m\{1 - (1 - n)\Sigma f_k(\sigma_m - \sigma_k)/[(1 - n)\sigma_m +$$

$$n\sigma_k]\}/\{1 + n\Sigma f_k(\sigma_m - \sigma_k)/[(1 - n)\sigma_m + n\sigma_k]\}. \qquad (20)$$

It is worthwhile to note that $\sigma_e$ is the ratio of the measured current to the measured electric field inside a sample. Because of this it does not depend on the shape of a sample, which is represented by the depolarization factor of a sample $N$, but naturally, depends on the shape of the grains, which is represented by the depolarization factor of the grains $n$. For the case of spherical grains we have:

$$\sigma_e = \sigma_m[1 - 2\Sigma f_k(\sigma_m - \sigma_k)/(2\sigma_m + \sigma_k)]/[1 + \Sigma f_k(\sigma_m - \sigma_k)/(2\sigma_m + \sigma_k)]. \qquad (21)$$

For a single sort of the inclusions in a matrix, we have:

$$\sigma_e = \sigma_m[2(1-f)\sigma_m + (1+2f)\sigma_i]/[(2+f)\sigma_m + (1-f)\sigma_i], \tag{22}$$

where $\sigma_i$ is the electric conductivity of the inclusions.

Let us consider now the case of a small difference between $\sigma_i$ and $\sigma_m$. That is let us consider a case when $\sigma_i = \sigma_m(1 + \delta_s)$, and $\delta_s \ll 1$. For this case we have:

$$\sigma_e/\sigma_m = 1 + f\delta_s - f(1-f)\delta_s^2/3 + \ldots . \tag{23}$$

For the case of arbitrary different values of $\sigma_i$ and $\sigma_m$, but for the small fraction of inclusions, $f - 1$, we have:

$$\sigma_e/\sigma_m = 1 - 3f[(\sigma_m - \sigma_i)/(2\sigma_m + \sigma_i)] + 3[f(\sigma_m - \sigma_i)/(2\sigma_m + \sigma_i)]^2 + \ldots . \tag{24}$$

Both limiting cases were considered in [2], and the results, represented by Eqs. (23) and (24) are in agreement with the ones, obtained in [2].

For the two-phased mixture $n = 1/3$, $\sigma_1 = \sigma$, $\sigma_2 = 0$, $f_1 = f$, and $f_2 = 1 - f$, we have:

$$\sigma_e = 2f\sigma\sigma_m/[2\sigma_m + (1-f)\sigma]. \tag{25}$$

Equations (22) and (25) do not show any singularity on $f$. So they cannot describe percolation phenomenon (see **Appendix C**).

The model, presented here, is relevant to the case of more or less homogeneous distribution of the different components of a mixture and it does not take into account the fluctuations inherent to the random distribution and clustering. That is why it could not be applied to the description of the percolation phenomena, which were studied completely and described in [3, 10, 11].

The concept of the homogeneous fields in each component of a fine mixture is quite natural and is obviously more accurate than the two simplest approaches, described in **Introduction**. Encouraging is the fact that in the limiting cases this simple model gives correct results and that the treatment of the experimental data obtained on the granular superconductive $YBa_2Cu_3O_7$ has shown the validity of the model also.

**6. Application to composite materials**

In **Section 5** Clausius-Mossotti approximation is applied to the modeling of effective electric conductivity of a sample, containing arbitrary amount of ellipsoidal conductive inclusions of the same shape and orientation in a conductive matrix. It was assumed that the current and the electric field in each inclusion and in the matrix are homogeneous ones, but they are different in different inclusions and in the matrix.

Now let us consider a sample with inclusions of one type [12]. Then Eq. (20) yields:

$$\sigma_e = \sigma_m[f\sigma + (1-n)(1-f)\sigma_m + n(1-f)\sigma]/[(1-n)\sigma_m + n\sigma - nf(\sigma - \sigma_m)]. \tag{26}$$

For volume fraction of the inclusions $f$ close to 1 and when the conductivity of the matrix $\sigma_m$ is essentially smaller than that of the inclusions $\sigma$, $(1-n)(1-f)\sigma_m$ can be neglected in the numerator of Eq. (26) comparative to $f\sigma$. The term $n(1-f)\sigma$ also could be neglected in the numerator of Eq. (26). In the denominator of Eq. (26) $\sigma_m$ can be neglected comparative to $\sigma$. After that we have [12]:

$$\sigma_e = f\sigma\sigma_m/[(1-n)\sigma_m + (1-f)n\sigma]. \tag{27}$$

When in the denominator of Eq. (27) $(1-n)\sigma_m$ can be neglected comparative to $(1-f)n\sigma$, Eq. (27) yields [12]:

$$\sigma_e = f\sigma_m/(1-f)n. \qquad (28)$$

It follows from Eq. (27) also that [12]

$$\sigma_m = (1-f)n\sigma\sigma_e/[f\sigma - (1-n)\sigma_e]. \qquad (29)$$

This equation can be used to calculate $\sigma_m$ from measured values of $\sigma, f, n$, and $\sigma_e$.

Let us regard a case of inclusions being conductive fibers of a shape of long cylinders of the length $l$ and diameter $d$. We assume that the volume fraction $f$ is close to 1. Depolarization factor of a long cylinder in the direction of the long axis can be approximated as one of a very long prolate spheroid with axes $l$, and $d$. For a very long prolate spheroid [13]

$$n = (d/l)^2[\ln(2l/d) - 1]. \qquad (30)$$

This quantity is an extremely small one. In the direction, perpendicular to the axis of a fiber, the depolarization factor is very close to 0.5.

From the symmetry of the regarded object follows that when the direction of the applied field is changed by $180°$, the direction of the measured current is also reverted. But the ratio of the measured field and current is not changed. From this follows that measured effective conductivity can be only an even function of applied electric field.

Let us calculate fields in the matrix and in the inclusions. Let $E_m$ denotes average field in the matrix and $E_i$ denotes that one in the inclusions. Average field in the sample, $\langle E \rangle$, is as follows:

$$\langle E \rangle = fE_i + (1-f)E_m. \qquad (31)$$

In the case of superconductive inclusions $E_i = 0$ and Eq. (31) yields [12]:

$$E_m = \langle E \rangle/(1-f). \qquad (32)$$

This field acts on the inclusions. When $f$ is close to unity, $E_m$ is essentially larger than $\langle E \rangle$.

To calculate the fields mentioned in Clausius-Mossotti approximation we need also following relationship:

$$\langle J \rangle = \sigma_e\langle E \rangle = f\sigma E_i + (1-f)\sigma_m E_m. \qquad (33)$$

It follows from Eqs. (31) and (33) that at any volume fraction $f$ of arbitrary value [12]

$$E_m = (\sigma\langle E \rangle - \langle J \rangle)/[(1-f)(\sigma - \sigma_m)], \qquad (34)$$

$$E_i = (\langle J \rangle - \sigma_m\langle E \rangle)/[f(\sigma - \sigma_m)]. \qquad (35)$$

From Eqs. (29), (34) and (35) follows [12] that

$$E_m = \{(\sigma - \sigma_e)[f\sigma - (1-n)\sigma_e]/\sigma(1-f)(f\sigma - \sigma_e + fn\sigma_e)\}\langle E \rangle, \qquad (36)$$

$$E_i = \{\sigma_e[f\sigma - (1-f)n\sigma - (1-n)\sigma_e]/f\sigma(f\sigma - \sigma_e + fn\sigma_e)\}\langle E \rangle. \qquad (37)$$

It should be noted, however, that Eqs. (36, 37) are applicable when volume fraction $f$ is close to unity.

As $E_m$ is the field, acting on the inclusions, the dependence $\sigma_m$ on electric field should be presented as $\sigma_m = \sigma_m(E_m)$.

Concentration of electric field in the vicinity of the ends of a nanofiber will be discussed in **Appendix D**.

**7. Spatial distribution of the magnetic field in a nonmagnetic matrix**

The magnetic field distribution in powder-in-nonmagnetic-matrix and granular samples depends on many factors: the shape of a sample, the shape of powder particles, the distribution of the powder particles in a sample, etc. The exact calculation of the field distribution in such a sample is a mathematical problem of extreme difficulties. So usually the problem is being addressed in some approximations. In [14] the regarded problem was addressed in Clausius-Mossotti approximation. This approximation is often used to calculate the distribution of the magnetic field **H** and the magnetic induction **B** in two-component mixture infinite media [1]. The essence of the approximation is to single out some volume $v$, inside a sample, which represents the inhomogeneous sample, and to regard the rest of the sample as some averaged substance. In such a way it is possible to calculate the ratio of the spatially averaged **B**, $\langle \mathbf{B} \rangle$, and **H**, $\langle \mathbf{H} \rangle$, i.e., the magnetic permeability $\mu = B/H$ (see, e.g. [1]).

Generalization of Clausius-Mossotti approach for samples of a finite volume was done in [8] for many-component mixtures. The essence of the approach remains the same: to regard inhomogeneity in a small representative volume $v$, and to regard a sample as a whole as consisting of some averaged media. According to this it was assumed that Eq. (4) is applicable.

Applying this to magnetic-powder-in-nonmagnetic-matrix and granular samples, equation for the space-averaged value of the magnetic field in a non-magnetic matrix was obtained [9],

$$\langle H_m \rangle = H_0 - 4\pi(N - n)\langle M \rangle, \qquad (38)$$

where $n$ is the depolarization factor of the magnetic particle (grain). For the randomly oriented particles $n$ is usually taken as 1/3 [6].

The value of the depolarization factor of a sample as a whole $N$ for inhomogeneous samples of a non-ellipsoidal shape is difficult to calculate because of the inhomogeneous distribution of the field inside a sample [7]. Strictly speaking $N$ value could be defined for ellipsoidal samples only [2]. So the simplest way to estimate actual $N$ value is to calculate it from the experimental data. For example, for the superconductive granular samples for this purpose it is possible to use the equation for $\langle M \rangle$ [9]:

$$\langle M \rangle = f(B_s - H_0)/4\pi[1 - fN - (1 - f)n], \qquad (39)$$

where $f$ is the volume fraction of a magnetic (superconductive) material and $B_s$ is the space-averaged remanent magnetic induction in a superconductive particle (grain). For $B_s = 0$ (low external fields) and for $n = 1/3$ Eq. (39) yields:

$$N = (1/3) + (2/3f) + (H_0/4\pi\langle M \rangle). \qquad (40)$$

Equation (40) should be used to calculate the actual $N$ values from measured values of $f$, $H_0$ and $\langle M \rangle$. As the magnetic field and the induction inside the magnetic particle (grain) [9],

$$H_g = H_0 - 4\pi[fN + (1-f)n]\langle M \rangle/f,$$

$$B_g = H_0 + 4\pi[1 - fN - (1-f)n]\langle M \rangle/f, \quad (41)$$

we have for the magnetic permeability of the magnetic particle (grain):

$$\mu_g = B_g/H_g = 1 + \{(fH_0/4\pi\langle M \rangle) - [fN + (1-f)n]\}^{-1}. \quad (42)$$

This equation could be used to calculate $\mu_g$ using experimental data on $f$, $H_0$, $\langle M \rangle$, $N$ and $n$.

Now let us consider the spatial distribution of the field around magnetic grains. Let us regard a simple model. We consider all the magnetic particles (grains) to be spheres of the same radius $Rf^{1/3}$, where $R$ is the radius of the representative sphere ($4\pi R^3/3 = v$ is a part of a volume of a sample, related to one magnetic grain).

According to the spirit of Clausius-Mossotti approach inhomogeneity is considered only within a representative sphere. Outside it the media is regarded to be a homogeneous one. Each magnetic particle (grain) of a volume $4\pi f R^3/3$ is a dipole with magnetic moment

$$m = 4\pi f R^3 \langle M \rangle/3f \equiv 4\pi R^3 \langle M \rangle/3, \quad (43)$$

which produces around it the space-averaged magnetic field, described by Eq. (38). So for $H_0$ parallel to the $z$-axis, $z$-component of the magnetic field in the range $Rf^{1/3} \le r \le R$ ($r$ is the distance from center of the sphere) is as follows:

$$H_{mz} = H_0 - 4\pi(N - n)\langle M \rangle - (m/r^3) + (3mz^2/r^5). \quad (44)$$

This field influences, e.g., the shape of the electronic paramagnetic resonance (EPR) line (EPR centers are usually distributed in a nonmagnetic matrix). Usually $H_0$ is the most significant term in Eq. (44). Keeping in mind to restrict our calculations of the line parameters, taking into account only linear (with respect to the magnetic moment) terms, let us restrict our consideration only with $z$-component of the field (two other components do not contribute to the linear terms). Thus Eq. (44) yields:

$$\langle (H_{mz} - \langle H_{mz} \rangle)^2 \rangle^{1/2} = 3.75 \langle M \rangle/f^{1/2},$$

$$\langle (H_{mz} - \langle H_{mz} \rangle)^3 \rangle^{1/3} = 2.56 \langle M \rangle/f^{2/3}. \quad (45)$$

For the low concentration of the magnetic powder $f$ more appropriate model is the model of the randomly distributed spheres. In this model each dipole produces its own field in the whole volume of a sample independently of the other dipoles (particles) [15]. In the framework of the model regarded we have:

$$\langle (H_{mz} - \langle H_{mz} \rangle)^2 \rangle^{1/2} = 3.75 \langle M \rangle [(1 + f)/f]^{1/2},$$

$$\langle (H_{mz} - \langle H_{mz} \rangle)^3 \rangle^{1/3} = 2.56 \langle M \rangle (1 + f + f^2)^{1/3}/f^{2/3}. \quad (46)$$

As the third moment of the line is not zero (see Eqs. (45) and (46)), the line is not symmetric. The field distribution influences the shape of EPR line, changing its virgin shape. Let the virgin shape be represented by some function $I(\omega, H_m)$ ($\omega$ is the angular frequency). Then due to the different values of $H_m$ in different points of a sample space the shape of the resonant line is changed and it could be calculated as a convolution, so the measured shape of the line is represented by the following relationship:

$$I_m(\omega, H_0) = (V/v) \int c(\mathbf{r}) I(\omega, H_m(\mathbf{r})) d\mathbf{r}, \qquad (47)$$

where $V$ is the sample volume ($V/v$ is the number of the magnetic grains in a sample), $c(\mathbf{r})$ is the concentration of the centers of the resonance in a non-magnetic matrix (inhomogeneous in general case), $I_m$ is the intensity per one center and the integration is taken over the representative volume $v$, in the case of the representative sphere model. The case of the random distribution of magnetic grains is to be considered separately.

As the third moment of the field distribution is not zero (see Eqs. (45) and (46)), the distribution of the field is not symmetric, so even in the case of a symmetric virgin line the observed line is expected to be asymmetric in some extent also.

## 8. Porous magnetic material

Let us consider a porous magnetic ellipsoidal sample in an external homogeneous magnetic field. Let us assume that we have pores of different types $k$. Each type is characterized by the depolarization factor of the pores $n_k$, and its volume fraction $f_k$. As it is assumed in Clausius-Mossotti approximation, the magnetic field in the matrix $H_m$ is supposed to be the homogeneous one and the magnetic field in each pore $H_k$ is supposed to be the homogeneous one also (both directed along the external homogeneous field). The magnetization of the matrix $M_m$ is assumed to be the homogeneous one and directed along the external field. In the case of a ferromagnetic matrix this means that the external magnetic field should be strong enough. In the case of a paramagnetic matrix $M_m = \chi H_m$ ($\chi$ is the magnetic susceptibility), which means that as far as $H_m$ is assumed to be homogeneous, $M_m$ should be considered homogeneous also. Averaged over the sample volume internal magnetic field can be expressed by the following relationship:

$$\langle H \rangle = (1 - f) H_m + \sum f_k H_k, \qquad (48)$$

where $f = \sum f_k$ is the total volume fraction of pores.

To calculate the field inside the pore we have to take into account that the field inside the magnetic matrix is $H_m$, and that to create the pore we should add to the pore volume a magnetization equal to $-M_m$. Hence, we have:

$$H_k = H_m + 4\pi n_k M_m. \qquad (49)$$

Taking into account that $\langle M \rangle = (1 - f) M_m$, Eqs. (4), (48) and (49) yield

$$H_m = H_0 - 4\pi N M_m + 4\pi M_m \sum f_k (N - n_k). \qquad (50)$$

Now let us calculate the magnetic energy of a system $E_m$. As it is well known, the change in the magnetic energy $dE_m$ due to the change in the magnetization $dM_m$ is described by the formula [16]:

$$dE_m = -H_m V_m dM_m, \qquad (51)$$

where $V_m$ is the volume of a matrix.

Equations (50) and (51) yield:

$$E_m/V_m = -M_m H_0 + 2\pi N M_m^2 + 2\pi M_m^2 \sum f_k (n_k - N). \qquad (52)$$

The magnetic energy of a ferromagnetic ellipsoid, containing one ellipsoidal pore, both ellipsoids oriented along the external homogeneous magnetic field, was calculated previously [15, 17 – 19]. Equation (52) represents a more general result, which for the case of a single pore of a small volume yields the same energy as calculated in [15, 17 – 19].

In polar materials in strong external fields, when moments are oriented along the external field, pores cause distortion of the lines of the field, leading to the increase in the energy of the field in the bulk of a sample. This leads to the elongation of the pores in the direction of a field, causing decrease in the energy of the field, when relaxation to a more equilibrium state is possible (see Fig. 2).

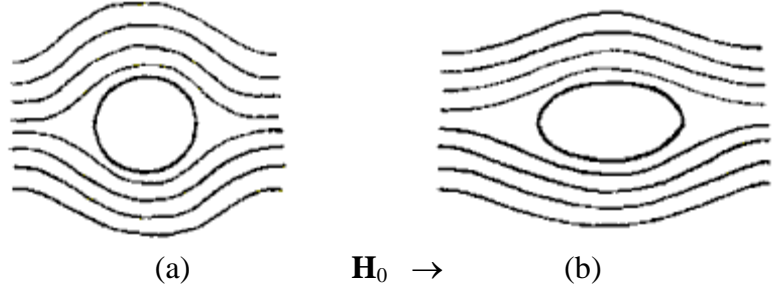

(a)    $\mathbf{H}_0 \to$    (b)

**Fig. 2.** Elongation of the spherical pore (a) causes decrease in the distortion of the magnetic field lines, which leads to a decrease in the magnetic energy of a system (b).

The equilibrium shape of a pore is the result of the competition between decreasing field energy and increasing surface energy when the pore becomes longer. As the size of the pore increases, the decrease in the energy of the field becomes so significant that it causes decrease in Gibbs free energy with the pore volume growth. This happens when the pore size becomes larger than a certain critical one, the typical value of which is about one micron. So larger pores tend to grow, leading to the cavitation phenomenon.

These phenomena were described by the author [15, 17 – 19] and could be observed in, e.g., alloys on the Co base as the Curie temperature $T_C$ of Co is 1400 K [20], and in this temperature range the diffusion processes are active. The kinetics and some properties of samples of the polar materials in strong external fields are discussed in [15, 17, 18].

**9. The equilibrium shape of a pore**

Let us consider a porous ferromagnetic ellipsoidal sample of the volume $V$ in an external homogeneous magnetic field $\mathbf{H_0}$. Let us assume that there are pores of different types $k$ in the sample. Each type is characterized by the depolarization factor $n_k$ and its volume fraction $f_k$. As it is assumed in Clausius-Mossotti approximation, the magnetic field in the matrix $H_m$ is assumed to be homogeneous one, and the magnetic fields in each pore $H_k$ are assumed to be homogeneous ones also, and all the fields are supposed to be directed along the external field (the external magnetic field is assumed to be strong enough to align all the magnetic moments along itself).

When only linear on the external pressure $P$ and the specific surface energy $\gamma$ terms are taken into account, the change in Gibbs free energy per unit volume of a sample (the specific Gibbs free energy) due to the formation of a pore $\Phi$ in Clausius-Mossotti approximation, as follows from Eq. (52), is presented by the following relationship:

$$\Phi = fP - 2\pi f(1-f)NM^2 + 2\pi(1-f)\langle n \rangle M^2 + \gamma \Sigma f_k(s_k/v_k), \qquad (53)$$

where $f = \Sigma f_k$, $N$ is the depolarization factor of a sample as a whole, $\langle n \rangle = \Sigma f_k n_k$, $n_k$, $s_k$, and $v_k$ are the depolarization factor, the surface area, and the volume of the *k*-th type pore, respectively.

More detailed analysis of various possible small contributions to Eq. (53) is given in [15, 21].

The equilibrium shape of the *k*-th type pores corresponds to the minimum of $\Phi$ on the eccentricity of the *k*-th type pores $\varepsilon_k$ and is determined by the following equation:

$$2\pi(1 - f)M^2(\partial n_k/\partial \varepsilon_k) + (\gamma/v_k)(\partial s_k/\partial \varepsilon_k) = 0. \tag{54}$$

For the case of the identical pores of a small volume fraction Eq. (54) yields [15, 17 – 19]:

$$\phi(x,\varepsilon) = \pi x^3 \{[(1 - \varepsilon^2)/2\varepsilon^3][\ln(1 + \varepsilon) - \ln(1 - \varepsilon) - 2\varepsilon] - N^*\} +$$
$$(9\pi/2)^{1/3} x^2 [(1 - \varepsilon^2)^{1/3} + (1 - \varepsilon^2)^{-1/6}(\varepsilon^{-1}\arcsin\varepsilon)] , \tag{55}$$

where the following dimensionless quantities were introduced [15, 17 – 19]:

$$\phi = 4M^4\Phi/\gamma^3; \quad x = 2M^2 v_p^{1/3}/\gamma; \quad N^* = N - (P/2\pi M^2), \tag{56}$$

where $v_p$ is the volume of one pore.

The dependence of the surface area of a spheroidal pore $s_p$ on the volume of a pore $v_p$ and the eccentricity $\varepsilon$, and the dependence of the depolarization factor $n$ on $\varepsilon$ were taken into account, deriving Eq. (55). There exists the equilibrium value of the eccentricity $\varepsilon_e$, which corresponds to the minimum of $\phi(x,\varepsilon)$ on $\varepsilon$. Equations (54) and (55) show that the equilibrium shape of a pore does not depend neither on the shape of a sample, nor on the external pressure. Expressions for $\varepsilon_e$ have been reported in [15, 17, 18] for a small nearly spherical pore, and for a large very extended pore. Using Eq. (56) these expressions can be rewritten as follows [19]:

$$\varepsilon_e = 0.99 x^{1/2} \text{ for } x \ll 1 \text{ , and for } x > 8.37$$

$$\varepsilon_e = 1 - 0.1(3.08/x)^{6/7}(\ln x/3.08)^{-6/7}[1 - (\ln x/3.08)]^{6/7}\{1 - (\ln x/3.08) - [\ln(\ln x)/3.08]\}^{-6/7} \tag{57}$$

The relative error in $\varepsilon_e$ in Eq. (57) is less than 4% [19]. The equilibrium eccentricity and the axes ratio (AR) of a pore of an intermediate size *x* were computed numerically [19]:

**Table 1.** The equilibrium eccentricity and AR as functions of *x* .

| *x* | 0.061 | 0.122 | 0.244 | 0.488 | 0.977 | 1.950 | 3.900 | 7.800 |
|---|---|---|---|---|---|---|---|---|
| $\varepsilon_e$ | 0.257 | 0.345 | 0.463 | 0.610 | 0.766 | 0.884 | 0.953 | 0.983 |
| AR | 1.035 | 1.065 | 1.128 | 1.262 | 1.556 | 2.139 | 3.301 | 5.294 |

Table 1 and Eq. (57) show that the elongation of the pore increases with the pore volume. Equation (54) shows that when the porosity of a sample is essential, the shape of each pore is determined by the factor $(1 - f)M^2$ [21] instead of just $M^2$, as it was for a single pore.

Anisotropy of a sample due to the presence of pores have been discussed and calculated in [17] (see also **Appendix A** in this paper). The rate of a pore volume change (bulk diffusion mechanism) was also discussed and calculated in [17, 18] (see also **Appendix B** in this paper).

## 10. Cavitation

Numerical calculations show that Gibbs free energy of the equilibrium pore $\phi_e = \phi(x,\varepsilon_e)$ at some positive $N^*$ increases from zero with the increase in $x$ and then reaches a maximum at some $x = x_c$ (the critical size of a pore). The position of this maximum depends on the value of $N^*$ only. With further increase in $x$ Gibbs free energy decreases, reaching zero at some $x = x_0$, and then decreases below zero with still further increase in $x$ (see Fig. 3).

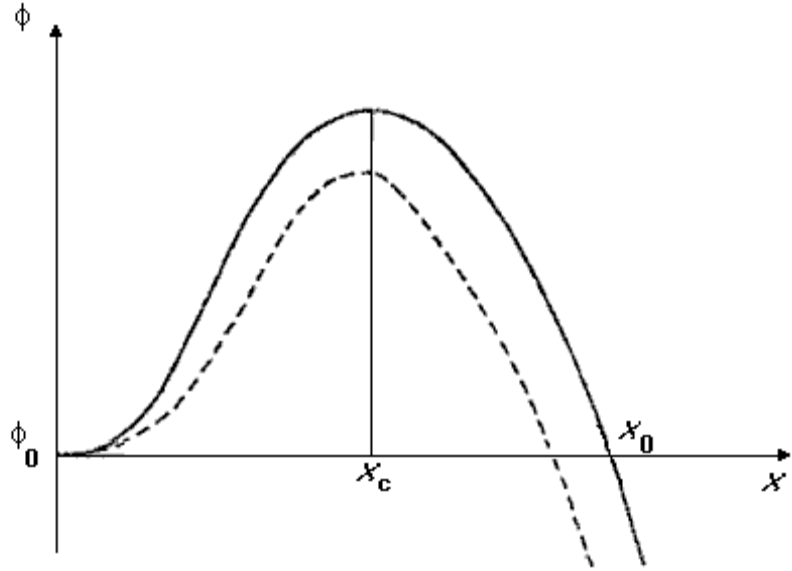

**Fig. 3.** The dependence of Gibbs free energy $\phi(x,\varepsilon_e)$ on the pore size $x$ has a maximum at some critical value of $x$, $x_c$. Pores larger than $x_c$ tend to grow. Solid line corresponds to the accepted model (invariable uniform moment **M** parallel to **H**$_0$). Dotted line takes into account partial relaxation of the direction of **M** near the pore. Because of some relaxation of **M** Gibbs free energy relaxes also and actual $x_c$ and $x_0$ are always smaller than the ones obtained from the model.

For $N^* = 1/3$, the computation yields $x_c = 5.16$ and $x_0 = 7.6$. As $N^*$ decreases $x_c$ and $x_0$ increase to infinity.

When the bulk diffusion is active, the volume a pore of a size smaller than the critical one, decreases. While the pores of a size larger than the critical one grow. This phenomenon is called cavitation. The rate of the diffusional change in the pore volume has been calculated in [17, 18].

**11. Ferromagnetic sample with ferromagnetic inclusions**

Let us consider a ferromagnetic ellipsoidal sample in external homogeneous magnetic field. Let us assume that there are ellipsoidal ferromagnetic inclusions of different types $k$ inside a sample. Each type of inclusions is characterized by the magnetization $M_k$, the depolarization factor of the inclusions $n_k$, and its volume fraction $f_k$. It is assumed in Clausius-Mossotti approximation that the magnetic field in a matrix $H_m$ is homogeneous one and the magnetic field in inclusions of each type $H_k$ is homogeneous also (both directed along the external homogeneous field). The magnetization of the matrix $M_m$ and the inclusions $M_k$ are assumed to be homogeneous, and directed along the external field. In the case of ferromagnetic materials this means that external magnetic field should be strong enough to result in parallel alignment of the moments.

Averaged over a sample volume internal magnetic field can be expressed by the following equation:

$$\langle H \rangle = (1 - f)H_m + \Sigma f_k H_k, \tag{58}$$

where $f = \sum f_k$ is the total volume fraction of inclusions.

To calculate the field inside the inclusion we have to take into account that the field inside the magnetic matrix is $H_m$ and that to create the magnetic moment of the inclusion we should add to the inclusion volume the magnetization equal to $M_k - M_m$. Hence, we have:

$$H_k = H_m + 4\pi n_k (M_m - M_k). \tag{59}$$

Taking into account that $\langle M \rangle = (1-f)M_m + \sum f_k M_k$, Eqs. (4), (58) and (59) yield:

$$H_m = H_0 - 4\pi(1-f)NM_m - 4\pi \sum f_k [n_k M_m + (N - n_k) M_k],$$

$$H_k = H_0 - 4\pi(1-f)NM_m + 4\pi n_k M_m - 4\pi n_k M_k - 4\pi \sum f_j [n_j M_m + (N - n_j) M_j]. \tag{60}$$

Now let us calculate the magnetic energy of the system $E_m$. As it is well known, the change in the magnetic energy $dE_m$ due to the change in the magnetization is described by [16]:

$$dE_m = -(1-f)VH_m dM_m - V \sum f_k H_k dM_k, \tag{61}$$

where $V$ is the volume of a sample. For the case of the same values of the magnetization of all the inclusions, when $M_k = M_0$, Eq. (61) can be integrated and using Eqs. (60) and (61) the total change in Gibbs free energy of a sample due to the formation of inclusions can be obtained:

$$\Phi/V = (\Phi_0/V) - \langle M \rangle H_0 + 2\pi N \langle M \rangle^2 + 2\pi(1-f)\langle n \rangle (M_m - M_0)^2 + \gamma \sum f_k s_k / v_k, \tag{62}$$

where $(\Phi_0/V)$ is the change in Gibbs free energy per unit volume due to the formation of a ferromagnetic phase, $\gamma$ is the specific surface energy, $s_k$ and $v_k$ are the surface area and the volume of the inclusion of a $k$-type, respectively, and

$$\langle M \rangle = fM_0 + (1-f)M_m, \text{ and } \langle n \rangle = \sum f_k n_k. \tag{63}$$

The Gibbs free energy of a ferromagnetic ellipsoid, containing one ellipsoidal pore, with both ellipsoids oriented along the external homogeneous magnetic field, was calculated in [17, 18]. Equation (62) represents a more general result, which for the case of a single pore of a small volume yields the same energy as calculated in [17, 18].

## 12. The equilibrium shape of inclusions

The shape of a ferromagnetic inclusion (of the critical size, in particular) is often far from spherical, and its energy is not exactly the same as that of a spherical inclusion. This is important for a detailed analysis of phase transformations. So it seems to be worthwhile to calculate the equilibrium shape of a ferromagnetic inclusion. As the depolarization factor and the surface area of the inclusion depend on its shape, Eqs. (62) and (63) show that $\Phi$ depends on the inclusion shape also. Here we regard ellipsoidal inclusions, whose shape is described by the eccentricity $\varepsilon_k$. The equilibrium eccentricity corresponds to the minimum of $\Phi$ and is described by the following equation:

$$2\pi(1-f)(M_m - M_0)^2 (\partial n_k / \partial \varepsilon_k) + \gamma \partial (s_k / v_k) / \partial \varepsilon_k = 0. \tag{64}$$

The equilibrium shape of a ferromagnetic inclusion is determined by the competition between the magnetic energy, which decreases as the inclusion elongates along the field, and the surface energy, which increases concomitantly. It could be achieved through the relaxation mechanisms like the surface and bulk diffusion in solid state.

## 13. Ceramic superconductors

In this section we shall deal with the magnetic properties of ceramic superconductors [9, 22]. We shall not take into account the difference in the depolarization factors of particles in the matrix. Anisotropy in physical properties of a material also will not be taken into account. So we shall consider an oversimplified picture of magnetic properties for sintered high temperature superconductors. But surprisingly enough this oversimplified model yields quite reasonable results.

As it is impossible to introduce the depolarization factor for a sample of an arbitrary shape [7, 23], we shall restrict ourselves by the shapes of ellipsoids as usual.

Equation (4) is the basis of the consideration. This equation is a well-known relation for the homogeneous samples. Application of Eq. (4) to the case of spatially inhomogeneous samples is the essence of the discussed model and it is logical to extend another well-known equation to our case:

$$\langle B \rangle = \langle H \rangle + 4\pi\langle M \rangle \equiv H_0 + 4\pi(1-N)\langle M \rangle. \tag{65}$$

Generally $\langle M \rangle$ could be represented as

$$\langle M \rangle = \langle M_{tr} \rangle + \langle M_{dia} \rangle, \tag{66}$$

where $\langle M_{tr} \rangle$ is the part of the measured magnetic moment, related to the trapped flux $\langle B_{tr} \rangle$, and $\langle M_{di} \rangle$ represents the diamagnetic response of a sample.

Trapping during the field-cooled process (see **Appendix C**) occurs at the irreversibility temperature $T_{ir}$, which is close to $T_c$, where the critical current $J_c$ (see **Apendix C**), is of a rather small value. Therefore, according to the critical state models, the trapped flux is distributed much more homogeneously compared to the case of the critical state at low temperature $T$ (the inhomogeneity of the flux distribution is proportional to $J_c$). In the case of small grains, the trapped flux could be regarded as being distributed almost homogeneously.

In the case of a homogeneous sample (e.g., single crystal) $\langle M_{tr} \rangle$ coincides with the so called remanent moment $\langle M_{rem} \rangle$, and $\langle M_{dia} \rangle$ represents the zero-field-cooled magnetization (see **Appendix C**), $\langle M_{zfc} \rangle$. For this case $\langle M \rangle = \langle M_{rem} \rangle + \langle M_{zfc} \rangle$, as was shown in [22, 24]. In the case of inhomogeneous sample, in particular granular one, the magnetic structure of a sample could be quite sensitive to rather low external fields [22]. This could strongly influence the dependence of $\langle M_{tr} \rangle$ and $\langle M_{dia} \rangle$ on the external applied field.

As the aim of this section is to describe the dependence of the measured magnetic moment on the magnetic structure of a sample, let us relate the measured magnetic moment $\langle M \rangle$ and the effective magnetic permeability of the inhomogeneous sample $\mu_e$. To do this we have to express the variations of the internal magnetic field $\Delta\langle H \rangle$ and the magnetic induction $\Delta\langle B \rangle$ caused by the presence of the external magnetic field $H_0$. As $\langle H \rangle$ and $\langle B \rangle$ are given by Eqs. (4) and (65) and at $H_0 = 0$, $\langle H \rangle = -4\pi N\langle M_{rem} \rangle$ and $\langle B \rangle = 4\pi(1-N)\langle M_{rem} \rangle$, we have:

$$\Delta\langle H \rangle = \langle H \rangle + 4\pi N\langle M_{rem} \rangle = H_0 - 4\pi N(\langle M_{tr} \rangle - \langle M_{rem} \rangle) - 4\pi N\langle M_{dia} \rangle,$$

$$\Delta\langle B \rangle = \langle B \rangle - 4\pi(1-N)\langle M_{rem} \rangle = H_0 + 4\pi(1-N)(\langle M_{tr} \rangle - \langle M_{rem} \rangle) + 4\pi(1-N)\langle M_{dia} \rangle. \tag{67}$$

To derive Eq. (67) we used Eq. (66) also. According to the definition of $\mu_e$,

$$\Delta\langle B\rangle \equiv \mu_e \Delta\langle H\rangle, \tag{68}$$

and Eqs. (66 - 68) we have:

$$\langle M\rangle = \langle M_{rem}\rangle - (H_0/4\pi)\{(1 - \mu_e)/[1 - N(1 - \mu_e)]\}. \tag{69}$$

The value of $\mu_e$ depends on the magnetic structure of a sample. When this structure does not depend on $H_0$, $M$ depends linearly on $H_0$. The magnetic structure is characterized in particular by the volume fraction of a superconductive material $f$. This value could be field-dependent because, e.g., in low field the volume of weak links contributes to it, while in higher fields weak links are no longer in the superconductive state, which leads to the decrease of the superconductive volume fraction. In the case of low field and random distribution of the superconductive fraction, magnetic induction could percolate through the sample only via nonsuperconductive components. This is possible only in the case when the volume fraction of a nonsuperconductive material $1 - f$ exceeds the percolation threshold. In this case $\mu_e$ is positive. Otherwise $\mu_e = 0$ and Eq. (69) acquires a form usual for the homogeneous superconductors.

When a sample contains the weak links, which surround the superconductive grains, and the external field is high enough for them to be in a normal state, we shall have the situation where the superconductive grains are separated and surrounded by a nonsuperconductive material. In this case, when the weak links are always percolating through the sample, the most adequate and available model for $\mu_e$ is Clausius-Mossotti one [25]. Now let us regard the concise description of the version of the model intended for the granular superconductors.

Let us consider a sample, which consists of superconductive inclusions surrounded by a nonsuperconductive volume. Now we consider one inclusion surrounded by a corresponding nonsuperconductive volume. In Clausius-Mossotti approximation the rest of the sample is regarded as being homogeneous. This means that the value of the magnetic field acting on the chosen inclusion could be evaluated by spreading the magnetic moment uniformly throughout the entire volume of a sample and compensating magnetic moment in a specific volume related to the chosen inclusion by adding to the volume the magnetic moment density of opposite sign $-\langle M\rangle$. Hence, the value of the magnetic field in the considered domain is:

$$H_m = \langle H\rangle - 4\pi n(-\langle M\rangle) = H_0 - 4\pi(N - n)\langle M\rangle, \tag{70}$$

where $n$ is the depolarization factor corresponding to the shape of a grain. In reaching Eq. (70) we have used Eq. (4).

As the density of the magnetic moment inside a superconductive inclusion $M_s = \langle M\rangle/f$, the magnetic induction inside the inclusion is described by the following equation:

$$B_s = \langle H\rangle + 4\pi(1 - n)M_s = H_0 - 4\pi(N - n)\langle M\rangle + 4\pi(1 - n)\langle M\rangle/f. \tag{71}$$

From this equation follows that

$$\langle M\rangle = f(B_s - H_0)/4\pi(1 - N_e), \quad N_e = fN + (1 - f)n, \tag{72}$$

where $N_e$ is the effective depolarization factor of a granular sample.

Equations (70) and (71) allow one to treat the experimental data properly (see [9, 22, 26] and Figs. 4 and 5). Figure 4(a) shows typical measured magnetization versus applied field curves for $\mathbf{H}_0$ parallel to the short axis of a sample ($c$) and to the long one ($r$) [22]. The depolarization factor in the $c$-direction is larger than that in the $r$-direction. This means that at the same value of the applied field the field value inside the sample is larger when the short axis is oriented along the field. The larger the internal field the larger the magnetization is, which is the case we see in Fig. 4(a). But when plotted as a function of effective magnetic field $H_e = H_0/(1 - N_e)$, magnetization in both directions at the same external fields practically coincide (see Fig. 4(b)). This occurs because the value of $H_e$ is respectively larger for the short direction at the same value of the external field $H_0$. In fact, this method could serve as an experimental tool to extract $N_e$ of polycrystalline samples.

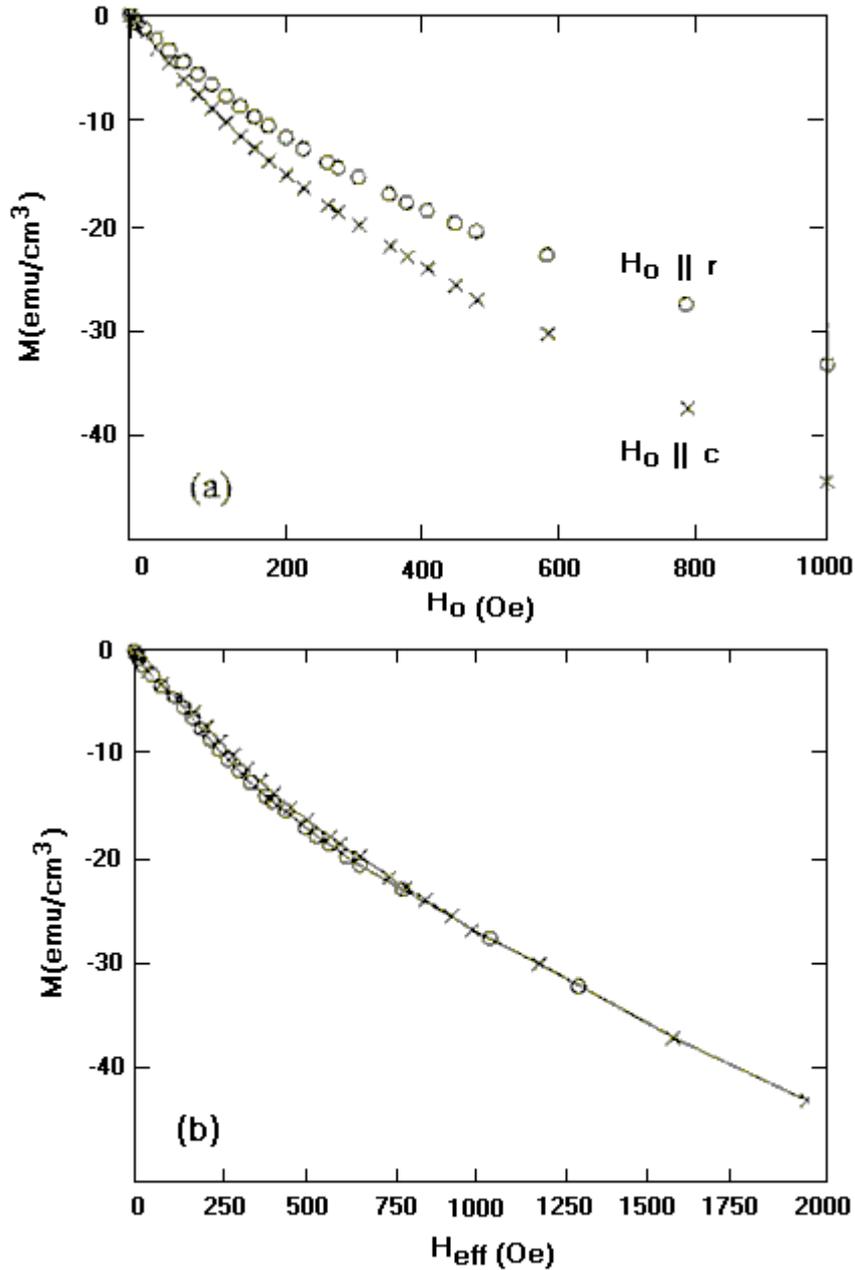

**Fig. 4.** Measured magnetization (points) at $\mathbf{H_0}$ parallel to the short (*c*) and long (*r*) axes of a sample, plotted (a) as a function of $H_0$ and (b) as a function of effective $H_e = H_0/(1 - N_e)$ [22]. When plotted as a function of $H_e$, measured magnetization in both directions at the same external fields practically coincide (b). This occurs because the value of $H_e$ is respectively larger for the short direction at the same value of the external field $H_0$.

The shielded fraction as a function of the external applied field $H_0$ was calculated from the experimental data according to the discussed model for several samples [22]. Results are presented in Fig. 5. Such approach could serve as experimental tool to extract experimental values of the critical fields of the weak links and the grains of polycrystalline samples.

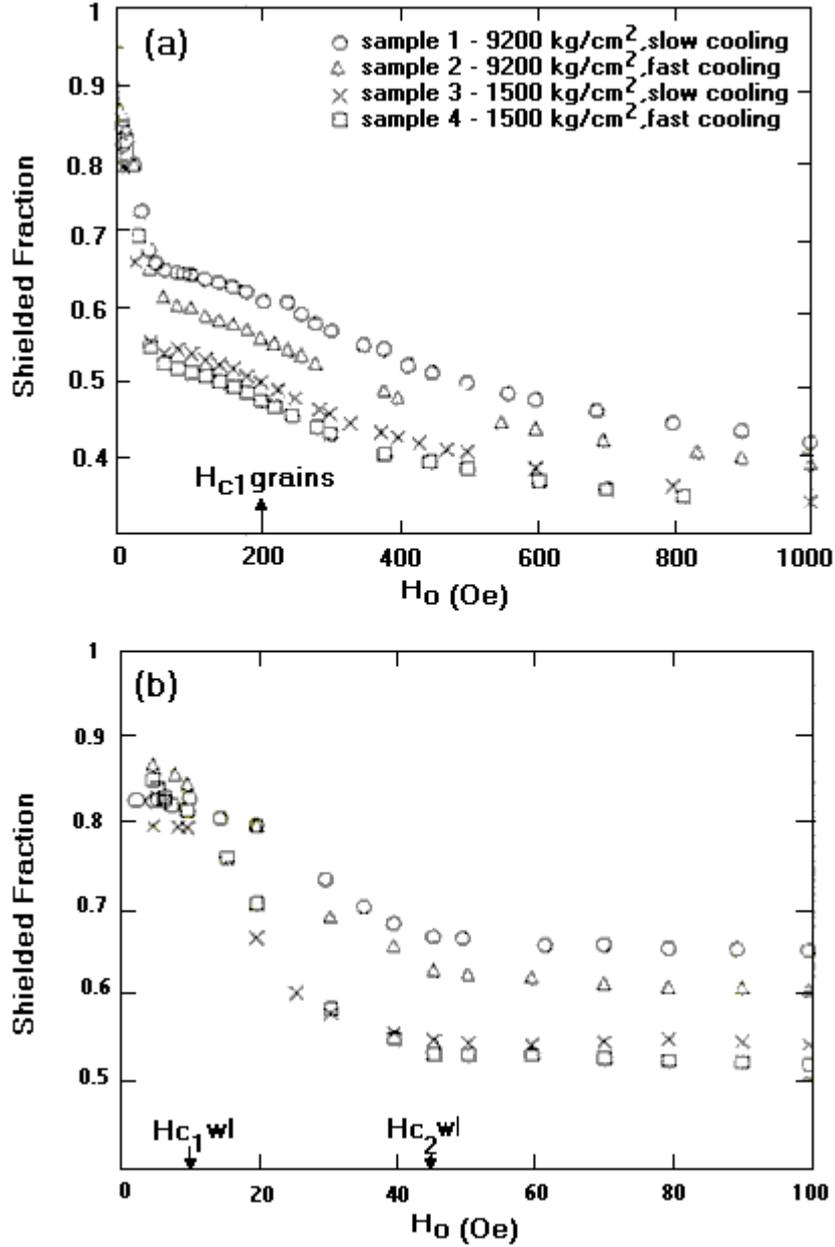

**Fig. 5.** The applied field dependence of the shielded fraction, calculated from the experimental data as $f = (1 - n)M/[M(N - n) - (H_0/4\pi)]$ [22]. Increase in the field leads to the decrease in the shielded fraction; (a) presents a wide range field picture, (b) gives an order of magnitude narrower interval of fields.

Now let us calculate the magnetic energy of a granular sample with a nonmagnetic matrix [26, 27]. For this purpose we should know the field inside the magnetic grain $H_g$. Taking into account Eq. (71) and that the induction inside the grain $B_g = H_g + 4\pi M_g \equiv H_g + 4\pi\langle M\rangle/f$ ($M_g$ is the grain magnetization) we have

$$H_g = H_0 - 4\pi N_e \langle M\rangle/f. \qquad (73)$$

So for the magnetic energy we have [27]:

$$dF_m = -H_g f V dM_g = -H_0 V d\langle M\rangle + 4\pi N_e(\langle M\rangle/f)V d\langle M\rangle,$$

$$F_m = -H_0 V \langle M \rangle + 2\pi N_e \langle M \rangle^2 V/f. \tag{74}$$

Equation (74) is applicable for the granular magnetic sample with a nonmagnetic matrix, including ceramic superconductive samples. In [26] it is used to calculate the value of the external field at which the magnetic induction starts penetrating through superconductive grains.

## 14. Conclusion

Described extension of Clausius-Mossotti model proved to be rather useful for applications to the inhomogeneous polar materials, in particular sintered ceramic materials [28 - 30]. In spite of the oversimplified nature of the regarded model, it proved to be useful for the interpretation of the experimental data in the field-cooled and zero-field-cooled superconductive ceramic samples [9, 22, 26]. Measured EPR data in non-magnetic matrices in superconductive-powder-in-polymer systems also show unexpectedly high accuracy of the reported simple approach [31]. So, Clausius-Mossotti approximation, which was started by Ottavanio Fabrizio Mossotti in 1846, still serves new data in polar materials.

Equilibrium shape of a pore in polar materials is investigated and the cavitation phenomenon is described. It appears that the larger the pore the more elongated it is in the given sample. Clausius-Mossotti model is applied also to calculate the effective electric conductivity of mixtures. The effective electric conductivity does not depend on the shape of a sample as a whole, as it is a coefficient between the current and the electric field inside a sample. This approach cannot serve the percolation phenomena, which is characteristic for the random distribution of the conductive and non-conductive particles, it is rather suitable for more or less ordered array of the particles of the components in a mixture. Measured magnetic moment, on the contrary, is sensitive to the shape of a sample, as it involves internal and external relationships.

## Appendix A. Pore contribution to the anisotropy

Anisotropic equilibrium pores may cause an additional anisotropy of various physical properties, such as some mechanical properties and the ferromagnetic anisotropy (difference in Gibbs free energies of a sample at different directions of a strong saturating external magnetic field) [17].

Let us consider the case when the equilibrium pore in a ferromagnet is formed at a sufficiently high temperature by the means of diffusion. If the pore shape remains unaltered during rapid cooling (quenching), the anisotropic shapes of the pore and the sample contribute to the energy of the magnetic anisotropy, and this contribution, according to Eq. (41), is as follows:

$$\Phi_a \equiv \Phi_\perp - \Phi = 2\pi f(1-f)(N - N_\perp)M^2 V + \pi(1-f)M^2(f - 3\langle n \rangle)V, \tag{75}$$

where $f = \Sigma f_k$ is the total volume fraction of pores, $N_\perp$ is the depolarization factor of a sample as a whole in the direction perpendicular to the longitudinal axis of a pore and $n$ is that of an axially symmetric pore. When all the pores are of one type $\langle n \rangle = fn$.

According to Eq. (75) even the spherical pore contributes to the anisotropy of a sample if a sample itself is anisotropic (the first term of the right-hand part of Eq. (75) remains non-zero). At $N = N_\perp$, $M^2 = 2 \times 10^6$ erg/cm$^3$, $n = 0.1$, $f = 0.01$, Eq. (75) yields $\Phi_a/V = 4.4 \times 10^4$ erg/cm$^3$. Hence, in some cases, such as thin films and sintered materials this contribution appears to be of a great importance.

## Appendix B. Diffusional kinetics of pores

Volume change rate of a small nearly spherical pore is obviously determined chiefly by the specific surface energy and external pressure. In this case ferromagnet demonstrates no specificity. Here let us consider a case when the surface diffusion is active enough for a pore to preserve permanently its equilibrium shape, and if the regarded pore is large enough, to be essentially elongated. For example, as was mentioned in **Section 10**, when $N^* = 1/3$, the computation yields $x_c = 5.16$ and $x_0 = 7.6$. As $N^*$ decreases, $x_c$ and $x_0$ increase to infinity.

When bulk diffusion mechanism is active, a pore of a size smaller than the critical one is being healed. While the pores of a size larger than the critical one, grow. The rate of a diffusional change in the pore volume of a long pore has been calculated [18]. In the absence of sources and sinks of vacancies in the bulk of a material

$$(dv_p/dt) = 12.37(Dv_a^2/\delta^{1/3}\ln\delta)[(2\pi M^2 N - P)/k_B T](v_c^{1/3} - v_p^{1/3}), \tag{76}$$

where $v_a$ is atomic volume, $v_c$ is the critical value of the pore volume, $D$ is a bulk self-diffusion coefficient, $k_B$ is Boltzman constant, $T$ is abolute temperature, and

$$\delta = (5x\ln 5x)^{-6/7} \leq 0.013. \tag{77}$$

When the density of the sources (sinks) of vacancies is sufficiently high, then

$$(dv_p/dt) = 3.38(Dv_a^2 v_p^{1/3}/\langle l \rangle \delta^{1/6})[(2\pi M^2 N - P)/k_B T](v_p^{1/3} - v_c^{1/3}), \tag{78}$$

where the mean free path of the excess vacancy, $\langle l \rangle$, is supposed to be much smaller than the smallest radius of the curvature of the pore.

Eqs. (76) and (78) are valid for $2\pi M^2 N > P$ only.

**Appendix C. Some concepts**

*Weak links*. It is well known that the magnetic field weakens the superconductivity. It starts penetrating the superconductive areas at lower critical field $H_{c1}$ and completely destroys the superconductivity at higher critical field $H_{c2}$ [32]. In inhomogeneous superconductors, like ceramic ones, $H_{c1}$ and $H_{c2}$ are also inhomogeneous. Areas with essentially low critical fields are called weak links. Usually they are situated near the boundaries between superconductive grains. With the increase in the magnetic field the weak links loose their superconductivity and contribute to the increase in the non-superconductive matrix volume, thus leading to the decrease in the shielded fraction.

*Depolarization factors*. It is well known that the magnetic field in a magnetic sample, subjected to a homogeneous external field $H_0$ is homogeneous only for the samples of the ellipsoidal shape [2]. For the ellipsoidal sample in the external homogeneous magnetic field directed along one of the main axes of the ellipsoid $k$, the magnetic field inside a sample $H_i$ is given by the following equation: $H_i = H_0 \sum 4\Sigma N_k M$ [15, 16]. This relationship was in some form assumed to be valid for the granular samples also in the framework of the extended Clausius-Mossotti model. In the latest relationship $M$ is the magnetization and $N_k$ is the depolarization factor of a sample with respect to the $k$-axis. It depends on the shape of the ellipsoidal sample only. There is a remarkable property of the values of the depolarization factors of a given ellipsoidal sample: $\Sigma N_k = 1$ [2]. As for a spherical sample all its axes are identical, all the three depolarization factors of the sample are equal to 1/3.

*Percolation*. The conductivity of a sample, which consists of a non-conductive matrix and random conductive inclusions, depends crucially on the filling factor, $f$. For low filling factors a sample is not conductive. When the value of the filling factor reaches some threshold $f_c$, bridging cluster is formed and the sample becomes conductive. So, at $f = f_c$ the singularity takes place and drastic change in the conductivity of a sample occurs. This phenomenon is called the percolation phenomenon [3, 10, 11].

*Field-cooled process*. This is the process of cooling a sample down to a superconductive state in the presence of the external magnetic field. As the magnetic field penetrates a non-superconductive sample and is expelled from a superconductive one, some of the field is trapped in a sample during the process, which leads to the formation of a remanent magnetic moment (see, e.g., [9]).

*Zero-field-cooled process*. This is a process of cooling a sample down to a superconductive state without the external magnetic field. After this process is performed no magnetic field is trapped in a sample and no remanent moment of a sample is found (see, e.g., [21]).

*Critical current.* The electric current weakens the superconductivity. The larger the current the larger the effect is. At some critical value of the current, the critical current, the superconductivity fails. The higher the temperature the lower the critical current is. The critical current turns to zero at the critical temperature $T_c$.

**Appendix D. Concentration of the electric field near the ends of a nanofiber**

When external electric field is applied along the fibers, the charges $\pm q$, giving rise to the electric field, which compensates that inside the conductive fibers, occur only at the fiber ends, where a well-known concentration of the electric field in the vicinity of sharp object takes place [2]. When the length of the fiber $l$ is much larger than its diameter $d$, the capacity of both fiber ends $C$ should be approximated as $d/\pi$ [33]. In [33] a fiber with flat ends was studied. Electrostatic potentials of the field produced by the charges $\pm q$ at the fiber ends could be estimated as $\varphi_{1,2} = \pm\pi q/\varepsilon d$ (here $\varepsilon$ the dielectric constant of the matrix), and the local electric field in the vicinity of the ends is $E_b = \pm 2\pi q/\varepsilon d^2$. Potential difference between the potentials on the two fiber ends could be approximated as $\delta\varphi = 2\pi q/\varepsilon d^2$. From the other hand, the value of this potential difference is $lE_m$, so that the total potentials at both ends are equal, and the current carriers of the fiber are in the equilibrium. As follows from this, $q = \varepsilon d l E_m/2\pi = \varepsilon d l \langle E\rangle/2\pi(1-f)$, and that the local concentrated field in the vicinity of the fiber ends is as follows [34]:

$$E_b = \pm l E_m/d = \pm l\langle E\rangle/d(1-f). \quad (79)$$

Here Eq. (32) was taken into account. When $f$ is comparable with unity, Eq. (79) is rather rough approximation.

Total electrostatic energy $U$ per unit volume of the fiber $v_f$ was calculated in [35]:

$$(U/v_f) = -\varepsilon[(l/d) + (1/\pi)][\langle E\rangle/\pi(1-f)]^2. \quad (80)$$

It is worthwhile to note that the electrostatic energy of the fiber is negative and contains a factor $(l/d)$, which is essentially larger than unity. It is also proportional to the dielectric constant of the matrix $\varepsilon$, and external applied electric field in square $\langle E\rangle^2$.